# Risk Measures for the Climate Transition Age:

# Why Symmetric Risk Budgets are unlikely to mitigate two-dimensionally asymmetric Climate Change risks

Lars Dijkstra[a]; Andreas G. F. Hoepner[bc]*; Fabiola I. Schneider[bcde]
& Agnieszka Słomka-Gołębiowska[fg]

[a] PGGM, Noordweg Noord 150, 3704 JG Zeist
[b] European Commission Platform on Sustainable Finance, Brussels, Belgium
[c] Michael Smurfit Graduate Business School & UCD Lochlann Quinn School of Business, University College Dublin, Carysfort Avenue, Blackrock, Co. Dublin, Ireland
[d] ESMA Sustainability Standing Committee, 201-203 Rue de Bercy, 75589 Paris
[e] Greenhouse Gas (GHG) Protocol Scope 3 Technical Working Group, c/WRI, 10 G Street NE, Suite 800, Washington DC, 20002, USA
[f] mBank S.A. is Ul. Prosta 18, 00-850 Warszawa, Poland
[g] Department of International Comparative Studies, Warsaw School of Economics (PL), Warsaw, Poland

## *Abstract*

*The most widely used investment risk measures - variance, standard deviation, and tracking error - treat upside deviations as equally undesirable as downside losses. In stable markets with assumed symmetric return distributions, this simplification is potentially harmless. In the climate transition age, it is not. Climate risk is two-dimensionally asymmetric, across assets and over time. Both dimensions are negatively skewed against investors. Hence, climate transition investment strategies deliberately deviate from benchmarks that reflect an insufficiently decarbonising global economy. When fiduciaries enforce tracking error budgets that penalise outperformance just as heavily as underperformance, they inadvertently constrain exactly the portfolio return opportunities that climate-aware investing seeks. We argue that replacing symmetric deviation-based risk measures with their downside-only counterparts - semi-variance, standard risk and trailing error - resolves this conflict without sacrificing any fiduciary discipline. The distinction is not merely semantic: as the gap between climate transition trajectories and market benchmarks widens over the coming decades, symmetric risk budgets will become binding constraints on transition strategies well before actual investment risk warrants it. We propose specific changes to mandate design that would allow institutional investors to pursue climate transition opportunities within genuinely risk-aware (rather than deviation-averse) frameworks.*

*** Corresponding Author:** andreas.hoepner@ucd.ie

**Keywords:** climate change mitigation, climate transition benchmarks, deviation, fiduciary duty, paris-aligned benchmarks, risk, standard risk,

**Acknowledgements:**
We are deeply grateful for invaluable insights and close collaboration from Michiel De Smet (National Bank of Belgium) and Pernille Jessen (AP Pension). We likewise substantially benefitted from feedback on concept or draft development provided by Hampus Adamsson, Markus Koch, Markus Nilsson, Antoine Picot, and Gabija Zdanceviciute. The views expressed in this manuscript do not necessarily the represent the views of the affiliated organisations.

## The Problem of Two-Dimensional Asymmetry: Climate Risk Is Negatively Skewed Across Assets and Over Time

The climate transition age is unprecedented. For the first time, the physical effects of climate change are intensifying while mitigation efforts remain incomplete, and major democracies and their parties show inconsistent commitment to the Paris Agreement. What makes climate risk distinctive from an investment measurement perspective is its fundamental asymmetry on both dimensions: it is negatively skewed across assets and over times.

At any given point in time, more assets are negatively affected by climate change than positive. Already in the next few years, global GDP losses of up to 2.8% and unemployment increases of up to 1.7 percentage may occur since global mean surface temperatures are approaching 1.5°C above pre-industrial levels, according to the Network for Greening the Financial System (NGFS).[1] In other words, physical damages and/or transition policy shocks pose material risks to the average financial asset since the number of companies physically damaged and/or curtailed by transition risks will substantially outweigh the number of companies benefitting from the damage rebuild and/or the transition legislations, respectively.

Across time, the average asset is more likely negative than a positive outlook with respect to climate change. Global temperature increases and the accumulation of heat energy in the earth system exhibit no mean reversion and no symmetry around a central tendency (Forster et al., 2024). The distribution of possible climate outcomes is heavily skewed toward adverse scenarios, with tipping points, such as a substantial slowdown in the Atlantic Meridional Overturning Circulation (AMOC), that could trigger abrupt, irreversible regime shifts (van Westen & Baatsen, 2025).

Unlike a pandemic or a non-nuclear war, where recovery restores prior conditions, many climate processes cross thresholds beyond which they cannot reverse on any policy-relevant timescale. In other words, while a pandemic or non-nuclear war are asymmetric across assets when they occur (i.e. cross-asset dimension), they are transitory instead of asymmetric over time, as a vaccine can be developed or peace can be found (i.e. time series dimension). Climate Change, however, is asymmetric across both dimensions – cross-asset & time series – until the climate transition is successfully completed with mankind and our planet jointly absorbing as much $CO_2e$ as we emit. Even at this point, which is very unlikely to occur before 2050, delays in physical effects may mean that the two-dimensional asymmetry of climate change risks continues for additional years or decades.

This two-dimensional asymmetry has a direct implication for portfolio risk measurement. Any framework that assumes symmetric return distributions, i.e. treats large gains as equally undesirable as large losses, will systematically mischaracterise the risk profile of climate-exposed assets. For climate-aware investors seeking to tilt portfolios toward the transition, the consequences are worse still: symmetric measures penalise the very outperformance that a successful transition strategy is designed to produce.

## The Measurement Problem: Deviation Is Not Risk

The conflation of deviation with risk has deep roots. When Harry Markowitz developed what became Modern Portfolio Theory, he devoted an entire chapter of his 1959 monograph to semi-variance - a measure that considers only returns falling below a target - precisely to avoid the false assumption that upside and downside deviations are equally undesirable. He was explicit about the trade-off: variance was computationally four-times cheaper on 1950s hardware, but semi-variance was the conceptually superior risk measure.[3]

Markowitz (1959) wrote that variance *"considers extremely high and extremely low returns equally undesirable. An analysis based on variance seeks to eliminate both extremes. An analysis based on semi-variance, on the other hand, concentrates on reducing losses."* He returned to this point in his 1990 Nobel Prize lecture[2], observing that *"semi-variance seems more plausible than variance as a measure of risk, since it is concerned only with adverse deviations."*

The limitations of mean-variance optimisation as adopted in practice extend beyond the symmetry problem. Kaplan and Idzorek (2024) argue that the investment industry has put "the cart before the horse" by focusing on single-period portfolio construction while neglecting the lifecycle finance framework that should precede it — a framework that accounts for human capital, liabilities, and consumption over time. Their observation that normative models should not be dismissed simply because investors frequently behave irrationally applies with equal force here: the widespread adoption of symmetric deviation measures does not make them correct, any more than the widespread neglect of lifecycle finance makes single-period optimisation sufficient. Our concern is narrower but complementary: even within the single-period framework, the industry's standard risk measures penalise outcomes that investors welcome (aka 'profits').

Writing from the practitioner side, Sortino and van der Meer (1991) early on argued that proper downside risk measurement "requires a clear distinction between uncertainty and risk," contending that standard deviation fails to capture what is actually at stake for investors. Knight (1921) and Keynes (1921) are usually credited for developing the distinction between risk (i.e. known unknowns) and uncertainty (i.e. unknown unknowns). The uncertainties around physical climate change are abnormally huge and two-dimensionally asymmetric with both dimensions skewed against investors. They present a very strong justification for applying the Precautionary Principle in any climate finance context.

Yet more than six decades after Markowitz's original work and more than three decades after these repeated warnings from both the theoretical and practitioner sides, the investment industry overwhelmingly relies on symmetric deviation measures. Standard deviation dominates absolute risk reporting. Tracking error, the standard deviation of active returns relative to a benchmark, is, as Sorensen, Alonso, and Belanger (2023) observe, the *"most ubiquitous restriction"* imposed on actively managed portfolios. Fund platforms, professional curricula, and regulatory frameworks almost universally present these deviation measures under the label of risk.

The practical consequence is significant. When a mandate specifies a tracking error budget of, say, 100 basis points, it constrains the portfolio's total deviation from the benchmark—upward and downward alike. In effect, what is labelled a risk budget functions as a deviation budget. This means that roughly half of the budget is consumed by observations that represent favourable outcomes for the investor. The actual risk measurement is diluted by approximately 50%.

Consider a concrete illustration in Figure 1. In weekly Russell 1000 return data from Q2 2023 through Q1 2025, the majority of observations are positive. Standard deviation assigns substantial risk scores to these gain weeks – four of its five highest risk identifications are in fact gains. The downside measure, by contrast, identifies the five largest losses as the five highest risk events. That is what a risk measure should do.

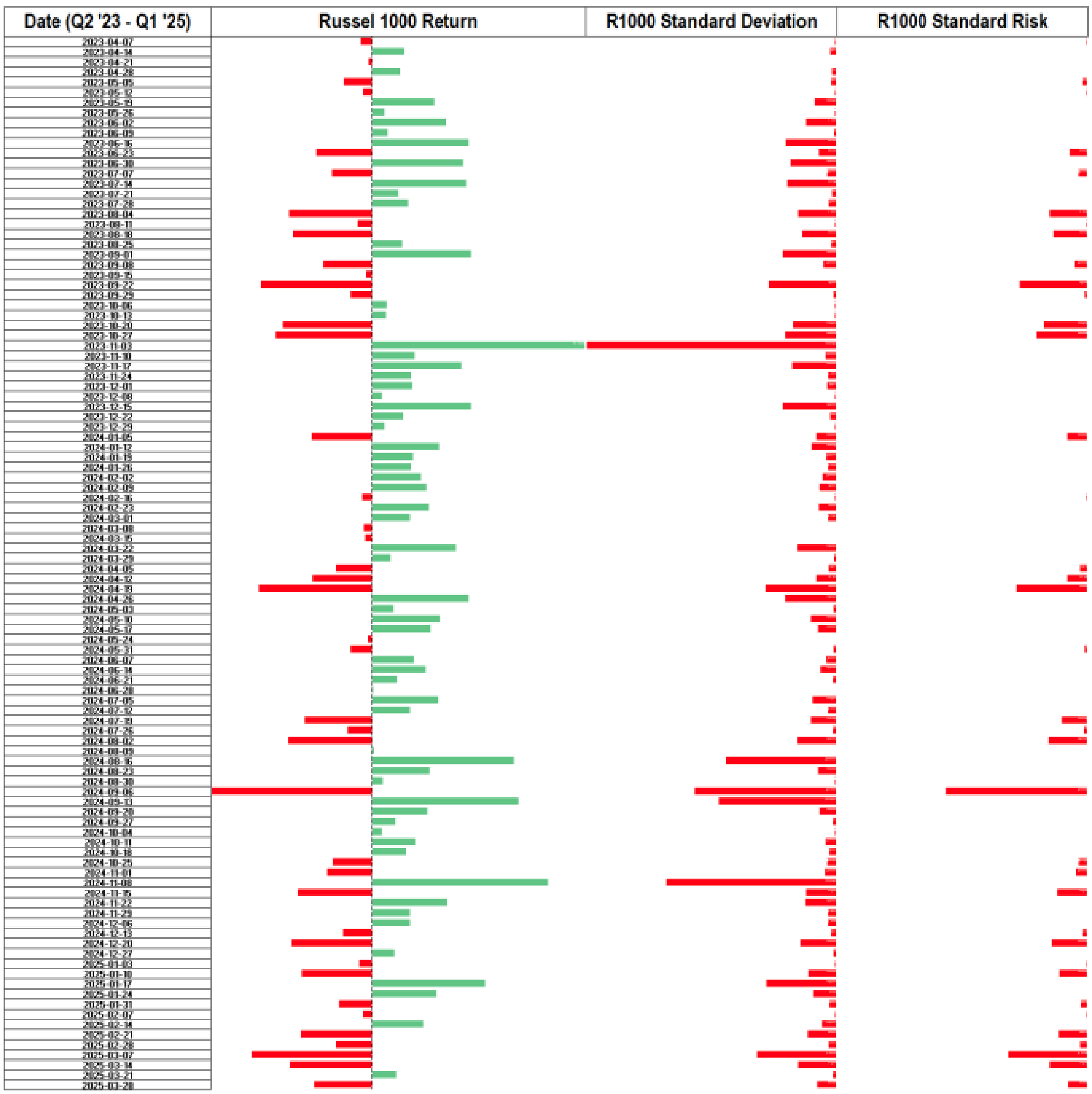

*Fig. 1. Comparison of Standard Deviation and Standard Risk identification of risk per week for the Russell 1000 for Q2 2023 to Q1 2025. Standard deviation's highest-risk identifications are*

*dominated by large gains; the downside measure correctly identifies large losses as the primary risk events.*

## The Climate-Specific Consequence: Tracking Error as a Leash on Transition

The distinction between deviation and risk becomes operationally critical when applied to climate transition investment strategies. These strategies deliberately tilt away from global market benchmarks because those benchmarks reflect the aggregate carbon exposure of an economy that is not decarbonising at the pace required by the Paris Agreement. If governments' commitments prove inconsistent, the gap between a climate transition portfolio's trajectory and the market benchmark's trajectory will widen over time.

Bolton, Kacperczyk, and Samama (2022) quantify the conflict between climate transition requirements and the ubiquitous use of tracking error restriction. They show that the common 1% threshold on tracking error would inevitably be broken in the early 2040s at the latest.

The constraint becomes binding not because the portfolio has taken excessive risk or because it underperformed the benchmark, but purely because it has deviated from a benchmark that is itself drifting further from Paris alignment. From a fiduciary perspective, this is a perverse outcome if a climate transition strategy produces superior returns. In this scenario, a climate transition strategy that outperforms its benchmark generates positive active returns - precisely the outcome the asset owner desires. Yet under a symmetric tracking error budget, that outperformance consumes risk budget in exactly the same proportion as equivalent underperformance would. The portfolio is, metaphorically, leashed to a benchmark that is heading in the wrong direction. The stronger its pull toward transition, the faster it exhausts its "risk" budget, even when the actual risk of loss has not increased.

Others have recently noticed elements of this problem. Practitioners at Fulcrum Asset Management have argued that *"climate alignment calls for bolder steps"* than conventional tracking error permits, in a paper titled "The tracking error error".[3] Invesco has asked whether one can *"satisfy* *climate objectives with low tracking error."*[4] We believe the root cause has not been fully identified in these discussions: the issue is not that tracking error budgets are too tight, but that tracking error itself is the wrong measure. It conflates deviation with risk. In any context where the investor's strategy is designed to produce directional outperformance, as climate transition strategies are, this conflation becomes actively harmful. The impact of this conflation worsens over time given the two-dimensional asymmetry of climate change risk.

## The Solution: Downside Risk Measures for Climate-Aware Mandates

The technical remedy is straightforward and has been available since Markowitz's original work. For every symmetric deviation measure in common use, there exists a downside-only counterpart that captures actual risk without penalising favourable outcomes:

| Symmetric Measure (Deviation) | Downside Counterpart (Risk) | What Changes |
|---|---|---|
| Variance | Semi-variance | Only returns below the mean are squared and averaged |
| Standard deviation | Standard risk | Only negative returns contribute to the measure |
| Tracking error | Trailing error | Only periods of benchmark underperformance are measured |

All three downside measures belong to the family of lower partial moments (LPMs), a general framework in which the investor specifies a target return and only shortfalls relative to that target are counted as risk. Semi-variance sets the target at the portfolio's mean return. Standard risk sets the target at zero. Trailing error[5] sets the target at the benchmark return. The mathematics is given in the Technical Appendix; the intuition is simple: an observation counts as risky if and only if the investor lost money or underperformed. Observations that made the investor better off are not risk.

When trailing error replaces tracking error in a climate-aware mandate, the constraint no longer penalises outperformance. A clima portfolio that beats its benchmark in a given period -because transition-positioned assets were rewarded - consumes zero trailing error budget for that period. Its full budget is reserved for periods of genuine underperformance. This roughly doubles the effective risk capacity available for climate transition opportunities.

## Practical Implications for Mandate Design

The shift from deviation-based to risk-based measurement is not merely a technical refinement. It carries concrete implications for how institutional investors structure their mandates, evaluate their managers, and fulfil their fiduciary obligations in the climate transition age. We propose four simple Principles for Risk Aware Investing to protect their returns from real risk and position their shareholder and stakeholder interest well in the age of the climate transition:

*Principle I:* We will incorporate accurate risk measurement (i.e. **focused on eliminating losses rather than deviations**) into investment analysis and decision-making processes.

*Principle II:* We will **seek accurate disclosure on risks** (i.e. focused on eliminating losses rather than deviations) by the entities in which we invest.

*Principle III:* We will be active, risk aware owners and incorporate accurate risk measurement **guidance into our ownership policies and practices**.

*Principle IV:* We will **promote acceptance and implementation of accurate risk measurement** (i.e. focused on eliminating losses rather than deviations) and the Principles of Risk Aware Investing within the investment industry.

## Conclusion

The argument we have made is narrow but consequential. We are not proposing a new climate risk model or a novel portfolio construction technique. We are pointing out that the risk measures most widely used in the investment industry - measures that were adopted for computational convenience in the 1950s - systematically mischaracterise the risk profile of any strategy designed to produce directional outperformance. In the age of two-dimensionally asymmetric climate risks, where climate transition portfolio strategies must increasingly diverge from market benchmarks, this mischaracterisation becomes a binding constraint on exactly the investment behaviour that the transition requires.

The fix is neither new nor controversial in principle. Markowitz himself advocated it. The downside risk literature has developed the tools thoroughly over the past several decades. What is needed now is adoption: a shift in industry practice from deviation budgets labelled as risk to genuine risk budgets that measure what investors actually worry about: the probability and magnitude of loss. Institutional investors who take the climate transition seriously cannot afford to treat this as a methodological nicety. When deviation budgets masquerade as risk budgets, fiduciary discipline and Paris alignment remain in unnecessary conflict.

## Technical Appendix: Downside Risk Measures

In the age of the climate crisis with its significant tail risks such as extreme weather, risk needs to be defined as the probability of a negative financial outcome. This negative outcome is that the index delivers less financial return than expected by the investor. Such expectation maybe the benchmark return (trailing error), a return above zero (standard risk) or a return exceeding its own average (semi-standard deviation). This means that an observation is considered risky if and only if it falls short of a set of financial expectations. Observations exceeding the expectations must not be considered a financial risk, since they instead represent an opportunity.

This definition is in line with the original writing of Markowitz (1959: 193-194), who explains that "[a]nalyses based on S[emi-variance] tend to produce better portfolios than those based on V[ariance]. Variance considers extremely high and extremely low returns equally undesirable. An analysis based on V[ariance] seeks to eliminate both extremes. An analysis based on S[emi-variance], on the other hand, concentrates on reducing losses." In other words, analyses based on variance seek to eliminate extremely high returns, which is clearly not in the interest of investors. Similarly, analyses based on tracking error seek to eliminate extremely high outperformance of the benchmark, which is likewise not in the interest of investors. Thus, an appropriate definition of risk must be used as the probability of negative financial outcome as it is applied in measures such as semi-variance, trailing error, value at risk or lower-partial moments.

To employ an accurate definition of risk which rates an observation is considered risky if and only if it falls short of a set of financial expectations, any of the following measures can be used.

| | |
|---|---|
| Lower Partial Moment (LPM) | $$LPM_{qp}(\psi) = \frac{1}{T-1}\sum_{t=1}^{T}\max\left[(\psi - r_{pt}),0\right]^{q}$$ |
| Semi-Variance (SV) | $$SV_{xp} = LPM_{2p}(\bar{r}_{xp})$$ |
| Semi-Standard Deviation (SSD) | $$SSD_{xp} = \sqrt{LPM_{2p}(\bar{r}_{xp})}$$ |
| Standard Risk | $$StandardRisk_p = \sqrt{LPM_{2p}(0)}$$ |
| Standard Skewed Risk | $$StandardSkewRisk_p = \sqrt{LPM_{3p}(0)}$$ |
| Shortfall Probability | $$ShortfallProbability_p = LPM_{0p}(0)$$ |
| Expected Average Loss | $$ExpectedLoss_p = LPM_{1p}(0)$$ |
| Expected Shortfall | $$ExpectedShortfall_p = LPM_{1p}(VaR\ Threshold)$$ |
| Trailing Error | $$TrailingError_p = \sqrt{LPM_{2p}(r_{xb})}$$ |

Notes: ψ is an investor's minimal acceptable return chosen by the individual investor, $r_{pt}$ is a the return (r) of portfolio (p) during observation interval (t), where the horizontal hat indicates the average where applicable. Excess returns above the risk free rate and benchmark are indicated by xp and xb, respectively. q indicates the level of investor risk aversion and is set to a value of 2 for semi-variance and semi-standard deviation. If q is set to a value of 3 in the LPM calculation, it creates a skewness focused version of LPM.

## Notes

1. NGFS, "Short-Term Climate Scenarios for Central Banks and Supervisors," 2025. Available at: https://www.ngfs.net/en/publications-and-statistics/publications/ngfs-short-term-climate-scenarios-central-banks-and-supervisors

2. Markowitz, H. Nobel Prize lecture, 1990. Available at: https://www.nobelprize.org/uploads/2018/06/markowitz-lecture.pdf. See also Markowitz (1991).

3. Fulcrum, 2024. "The tracking error error: why climate alignment calls for bolder steps". Available at: https://fulcrumasset.com/wp-content/uploads/2024/07/fulcrum-the-tracking-error-error-why-climate-alignment-calls-for-bolder-steps.pdf

4. Invesco, 2024. "Can You Satisfy Climate Objectives with Low Tracking Error?". Available at: https://www.invesco.com/nl/en/insights/satisfy-climate-objectives-with-low-tracking-error.html

5. See p.34 for an in-depth definition of trailing error https://finance.ec.europa.eu/document/download/50a36d31-606a-4f3b-ac49-710768162cef_en?filename=241219-sustainable-finance-platform-proposal-transition-benchmarks_en.pdf

6. ATP, 2023. Risk and Financial Condition Report. Available at: https://www.atp.dk/en/dokument/risk-and-financial-condition-report-2023. See also Man Group, "Covering Your Tail: Expected Shortfall," 2023. Available at: https://www.man.com/insights/covering-your-tail-expected-shortfall